# On the Metric of Quantum States with Lorentzian Signature

**Aalok Pandya***

Department of Physics, University of Rajasthan, Jaipur 302004, India.

In this note we comment on yet another way of describing the metric of quantum states with Lorentzian signature. For this, we consider the metric of quantum states and make successive transformations, exploiting relationship between $S^3$ and $SU(2)$.

______________________________

* *E-mail*: aalok_cosmos@yahoo.com

The problem of formulating metric of quantum states in the configuration space with the signature (+, +, +, -), has been addressed by Aalok et al [1]. However, the metric of quantum state space had been formulated in various attempts [2-4], but the metric of quantum state space is always positive definite as it is metric in the ray space [2]. In this note, we comment on yet another way of describing metric of quantum state but with signature (+, +, +, -). For this, we consider the following line element in the Minkowski space: $ds^2 = dx^2 + dy^2 + dz^2 - c^2 dt^2$. (1)

Which is a Lorentzian or a pseudo-Riemannian metric, and it can also be written as: $ds^2 = dx_1^2 + dx_2^2 + dx_3^2 + dx_4^2$; $x_1 = x, x_2 = y, x_3 = z, x_4 = ict$. (2)

Now we reparametrize this metric with the following consideration: instead of using the ordinary polar coordinates, we exploit the relationship between $S^3$ and $SU(2)$, with following complex coordinates [2],

$$Z_1 = x_1 + ix_2 = r\cos\frac{\theta}{2}\exp(\chi + \varphi);$$

$$Z_2 = x_3 + ix_4 = r\sin\frac{\theta}{2}\exp(\chi - \varphi);$$

with $0 \leq \theta < \pi, 0 \leq \varphi < 2\pi, 0 \leq \chi < 4\pi$. (3)

With these coordinates the above line element transform as;

$$ds^2 = dx^2 + dy^2 + dz^2 - c^2 dt^2$$
$$= dZ_1 d\bar{Z}_1 + dZ_2 d\bar{Z}_2. \qquad (4)$$

Which is in the Minkowski space: (+, +, +, -), as described by Eguchi and Hanson [5]. Now we recall the metric of quantum evolution, as proposed by Anandan [2]: $ds^2 = 2g_{\bar{\mu}\nu}d\bar{Z}^\mu dZ^\nu$ ; (5)

where $g_{\mu\nu} = \left( \left\langle \frac{\partial \tilde{\Psi}}{\partial Z^\mu} \Big| \frac{\partial \tilde{\Psi}}{\partial Z^\nu} \right\rangle - \left\langle \frac{\partial \tilde{\Psi}}{\partial Z^\mu} \Big| \tilde{\Psi} \right\rangle \left\langle \tilde{\Psi} \Big| \frac{\partial \tilde{\Psi}}{\partial Z^\nu} \right\rangle \right)$; (6)

and $Z^\mu$ represent complex coordinates.

Now, with $Z_1 = x_1 + ix_2$ and $Z_2 = x_3 + ix_4$ ; we reformulate the metric and the line element given by (5) and (6).

For that, we take $\frac{\partial}{\partial Z_1} = \frac{\partial}{\partial x_1} + i\frac{\partial}{\partial x_2}$ and $\frac{\partial}{\partial \bar{Z}_1} = \frac{\partial}{\partial x_1} - i\frac{\partial}{\partial x_2}$,

also $\frac{\partial}{\partial Z_2} = \frac{\partial}{\partial x_3} + i\frac{\partial}{\partial x_4}$ and $\frac{\partial}{\partial \bar{Z}_2} = \frac{\partial}{\partial x_3} - i\frac{\partial}{\partial x_4}$. (7)

With $dZ_1 = dx_1 + idx_2$ and $d\bar{Z}_1 = dx_1 - idx_2$;

$dZ_2 = dx_3 + idx_4$ and $d\bar{Z}_2 = dx_3 - idx_4$. (8)

Thus using (7) and (8) in (6); we get a diagonal metric form

$ds^2 = 2g_{11}d\bar{Z}_1 dZ_1 + 2g_{22}d\bar{Z}_2 dZ_2$,

i.e. $ds^2 = 2g_{11}(dx^2 + dy^2) + 2g_{22}(dz^2 - c^2 dt^2)$. (9)

This implies $g_{11} = \left( \left\langle \frac{\partial \Psi}{\partial x} - i\frac{\partial \Psi}{\partial y} \Big| \frac{\partial \Psi}{\partial x} + i\frac{\partial \Psi}{\partial y} \right\rangle - \left\langle \frac{\partial \Psi}{\partial x} - i\frac{\partial \Psi}{\partial y} \Big| \Psi \right\rangle \left\langle \Psi \Big| \frac{\partial \Psi}{\partial x} + i\frac{\partial \Psi}{\partial y} \right\rangle \right)$,

and $g_{22} = \left( \left\langle \frac{\partial \Psi}{\partial z} - i\frac{\partial \Psi}{c\partial t} \Big| \frac{\partial \Psi}{\partial z} + i\frac{\partial \Psi}{c\partial t} \right\rangle - \left\langle \frac{\partial \Psi}{\partial z} - i\frac{\partial \Psi}{c\partial t} \Big| \Psi \right\rangle \left\langle \Psi \Big| \frac{\partial \Psi}{\partial z} + i\frac{\partial \Psi}{c\partial t} \right\rangle \right)$. (10)

Thus, if we calculate four metric elements $\eta_{11}, \eta_{22}, \eta_{33}, \eta_{44}$, we find:

$$\eta_{11} = \eta_{22} = g_{11} \text{ and } \eta_{33} = \eta_{44} = g_{22}. \tag{11}$$

Here, $\eta_{11}$ is the coefficient of $dx^2$ in the expression of $ds^2$, and similarly $\eta_{22}$ is coefficient of $dy^2$, $\eta_{33}$ is coefficient of $dz^2$ and $\eta_{44}$ is coefficient of $dt^2$.

*Remarks*:

(1) Though, we had a positive definite metric on one hand, we ended up with a metric with the signature (+, +, +, -). We notice that this could happen as a result of the transformations into the specifically chosen space-time, and in the expression

$$ds^2 = 2g_{11}(d\bar{Z}_1 dZ_1) + 2g_{22}(d\bar{Z}_2 dZ_2) = 2\eta_{11} dx^2 + 2\eta_{22} dy^2 + 2\eta_{33} dz^2 - 2\eta_{44} c^2 dt^2,$$

the signature of the space had already been decided by

$$(d\bar{Z}_1 dZ_1) \text{ and } (d\bar{Z}_2 dZ_2). \tag{12}$$

(2) The curvature of this metric vanishes as the metric coefficients are under integrals and therefore all their derivatives are zero.

**Summarily, in equation (12) we have reformulated the metric of quantum states with the Lorentzian signature. In other words, we have transformed the metric of the quantum state space for $SU(2)$ on $S^3$ with the signature of Minkowski space.**